\documentclass[onecolumn,prl,aps]{revtex4}
\usepackage{amssymb}
\usepackage{epsfig,amsmath}
\usepackage{graphicx}
\usepackage{dcolumn}
\usepackage{bm}

\newcommand{\beq}{\begin{equation}}
\newcommand{\eeq}{\end{equation}}
\newcommand{\beqa}{\begin{eqnarray}}
\newcommand{\eeqa}{\end{eqnarray}} 
 
 \newcommand{\ka}{\kappa}

\newcommand{\la}{\langle}
\newcommand{\ra}{\rangle}

\def\ajp#1{{ Am.\ J.\ Phys.} {\bf #1}}

\def\npho#1{{Nature\ Phot.} {\bf#1}}
\def\njp#1{{ New\ J.\ Phys.} {\bf#1}}

\def\oc#1{{ Opt.\ Commun.} {\bf#1}}
\def\ol#1{{ Opt.\ Lett.} {\bf#1}}
\def\pla#1{{ Phys.\ Lett. A\/} {\bf#1}}

\def\pra#1{{ Phys.\ Rev. A\/} {\bf#1}}

\def\pre#1{{ Phys.\ Rev. E\/} {\bf#1}}
\def\prl#1{{ Phys.\ Rev.\ Lett.} {\bf#1}}

\def\sci#1{{ Science} {\bf#1}}

\begin{document}

\title{Shifting the Quantum-Classical Boundary:\\ Theory and
Experiment for Statistically Classical Optical Fields}
\author{Xiao-Feng Qian}
\email{xfqian@pas.rochester.edu}
\author{Bethany Little}
\author{John C. Howell}
\author{J.H. Eberly}

\affiliation{Center for Coherence and Quantum Optics and the Department of
Physics \&
Astronomy\\
University of Rochester, Rochester, New York 14627}

\begin{abstract}
The growing recognition that entanglement is not exclusively a quantum property, and does not even originate with Schr\"odinger's famous remark
about it [Proc. Camb. Phil. Soc. {\bf 31}, 555 (1935)], prompts examination of its role in marking the quantum-classical boundary. We have done
this by subjecting correlations of classical optical fields to new Bell-analysis experiments, and report here values of the Bell parameter greater
than ${\cal B} = 2.54$. This is many standard deviations outside the limit ${\cal B} = 2$ established by the Clauser-Horne-Shimony-Holt (CHSH) Bell
inequality [Phys. Rev. Lett. {\bf 23}, 880 (1969)], in agreement with our theoretical classical prediction, and not far from the Tsirelson limit ${\cal B} = 2.828...$. These results cast a new light on the standard quantum-classical boundary description, and suggest a reinterpretation of it.
\end{abstract}

\maketitle

\noindent{\bf Introduction:} For many decades the term ``entanglement" has been attached to the world of quantum mechanics \cite{Schr}. However, it is true that non-quantum optical entanglement can exist (realized very early by Spreeuw \cite{Spreeuw}) and its applications have concrete consequences. These are based on entanglements between two, or more than two, degrees of freedom, which are easily avalable classically \cite{Spreeuw, Ghose-Samal01, Lee-ThomasPRL, Borges-etal, Goldin-etal}. Multi-entanglements of the same kind are also being explored quantum mechanically \cite{Tiranov-etal}. Applications in the classical domain have included, for example, resolution of a long-standing issue concerning Mueller matrices \cite{Simon-etal}, an alternative interpretation of the degree of polarization \cite{Qian-EberlyOL}, introduction of the Bell measure as a new index of coherence in optics \cite{Kagalwala-etal}, and innovations in polarization metrology \cite{Toppel-etal}.  Here we present theoretical and experimental results extending these results by showing that probabilistic classical optical fields can exhibit violations of the Clauser-Horne-Shimony-Holt (CHSH) Bell inequality \cite{CHSH} of quantum strength. This is evidence of a new kind that asks for reconsideration of the common understanding that Bell violation signals quantum physics. We emphasize that our discussion focuses on non-quantum entanglement of non-deterministic classical optical fields, and does not engage issues such as non-locality that are important for some applications in quantum information.

The observations and applications of non-quantum wave entanglement noted above \cite{Simon-etal, Qian-EberlyOL, Kagalwala-etal, Toppel-etal, Spreeuw, Ghose-Samal01, Lee-ThomasPRL, Borges-etal, Goldin-etal} exploited non-separable correlations among two or more modes or degrees of freedom (DOF) of optical wave fields. Nonseparable correlations among modes are an example of entanglement \cite{vanEnk}, but are  not enough for our present purpose. In addition, we want to conform to three criteria that Shimony has identified for Bell tests \cite{Shimony}, facts of quantum Nature that must be satisfied when examining possible tests of the quantum-classical border. Fortuitously, the ergodic stochastic optical fields of the classical theory of partial coherence and partial polarization (see Wolf \cite{Wolf59}) satisfy these criteria fully (see Suppl. Materials \cite{Suppl.Mat.}), and we have used such fields as our test bed.

\noindent{\bf Background Theory:} We will deal here only with the simplest suitable example, the theory of completely unpolarized classical light, and have explained elsewhere (see \cite{Suppl.Mat.}, \cite{Qian-Eberly-arX13}) the generalizations needed to treat partially polarized fields, which lead to the same conclusions. In all cases there are only two degrees of freedom (DOF) to deal with, namely the direction of polarization and the temporal amplitude of the optical field. In both classical and quantum theories these are fundamentally independent attributes. An electric field, for a beam travelling in the $z$ direction, is written
\beq
\label{Efield}
\vec{E}(t) = \hat{x}E_{x}(t) + \hat{y}E_{y}(t).
\eeq
In the classical theory of unpolarized light \cite{{Brosseau-Wolf}} the
optical field's two amplitudes $E_x$ and $E_y$ are statistically completely
uncorrelated, and are treated as vectors in a stochastic function space.
A scalar product of the vectors in this space corresponds physically to
observable correlation functions such as $\la E_xE_y\ra$. For
unpolarized light we have $\la E_xE_x\ra = \la E_yE_y\ra$, and $\la
E_xE_y\ra = 0$.

Now it is possible to talk of entanglement of the classical field. This
is because entangled states are superpositions of products of vectors
from different vector spaces, whenever the superpositions can't be rearranged
into a single product that separates the two spaces \cite{Schr}. Looking again at
(\ref{Efield}), we see that this is the case because we've taken $\vec
E$ to be unpolarized. That is, by the definition of unpolarized light,
there is no direction $\hat u$ of polarization that captures the total
intensity, so $\vec E(t)$ can't, for any direction $\hat u$, be written
in the form $\vec E(t) = \hat u F(t)$, which would factorably separate the
polarization and amplitude DOF \cite{circpol}.

Beyond its probabilistic indeterminacy, the $\vec E$ in (\ref{Efield})
has other quantum-like attributes -- it has the same form as a quantum
state superposition and can be called a pure state in the same sense,
more precisely a two-party state living in two vector spaces at once,
polarization space for $\hat x$ and $\hat y$, and infinitely continuous
stochastic function space for $E_x$ and $E_y$.

The Bell inequality most commonly used for correlation
tests is due to Clauser, Horne, Shimony and Holt (CHSH) \cite{CHSH}. It
deals with correlations between two different DOF when each is
two-dimensional. The Schmidt Theorem of analytic function theory
\cite{Schmidt} ensures two-dimensionality, by guaranteeing that among
the infinitely many dimensions available to the amplitudes in
(\ref{Efield}), only two dimensions are active. This is a
consequence arising just from the fact that the partner polarization
vectors $\hat{x}$ and $\hat{y}$ live in a two-dimensional space.

For convenience, we introduce $\vec e$, the field normalized to the
intensity $I = \la E_xE_x +  E_yE_y \ra$:
\beq
\label{efield}
\vec{e}(t) \equiv \vec E(t)/\sqrt I = \{\hat{x}e_{x}(t) + \hat{y}e_{y}(t)\},
\eeq
where now $\la \vec e \cdot \vec e \ra = \la e_xe_x + e_ye_y\ra = 1$.

For some simplification in writing, we will use Dirac notation for the
vectors without, of course, imparting any quantum character to the fields.
The unit polarization vectors $\hat x$ and $\hat y$ will be renamed as
$\hat x \to |u_1\ra$ and $\hat y \to |u_2\ra$ and the unit  amplitudes
will be rewritten $e_x \to |f_1\ra$ and $e_y \to |f_2\ra$. If desired, the
Dirac notation can be discarded at any point and the vector signs and hats
re-installed. For the case of unpolarized light we have $\la u_1| u_2 \ra
= 0$ and $\la f_1| f_2 \ra = 0$. Unit projectors in the two spaces take
the form $1 = |u_1\ra\la u_1| + |u_2\ra\la u_2|$ and $1 = |f_1\ra\la f_1|
+ |f_2\ra\la f_2|$. In this notation, and in the original notation for
comparison, the field takes the form
\beqa
\label{edef}
\vec E/\sqrt{I} &=&  \hat x\ e_x + \hat y\ e_y \nonumber \\
&=& |{\bf e}\ra = \Big(|u_1\ra |f_1\ra + |u_2\ra |f_2\ra\Big)/\sqrt{2}.
\eeqa
In this notation the field actually looks like what it is, a  two-party
superposition of products in independent vector spaces, i.e., an entangled
two-party state (actually a Bell state). Here the two parties are the
independent polarization and amplitude DOF.

The notation for a CHSH correlation coefficient $C(a,b)$ implies that
arbitrary rotations of the unit vectors $| u_j \ra$ and $| f_k \ra (j,k =
1,2)$ through angles $a$ and $b$ can be managed independently in the two
spaces. An arbitrary rotation through angle $a$ of the polarization
vectors $|u_1\ra$ and $|u_2\ra$ takes the form
\beqa \label{uRotation}
&&|{u}_{1}^{a}\rangle =\cos a|{u}_{1}\rangle -\sin a|{u}_{2}\ra \quad
{\rm and} \notag \\
&& |{u}_{2}^{a}\ra =\sin a|{u}_{1}\ra +\cos a|{u}_{2}\ra.
\eeqa
For function space rotations we have
$|f_{1}^{b}\ra$ and $|f_{2}^{b}\ra$ defined similarly:
\beqa \label{fRotation}
&&|{f}_{1}^{b}\rangle =\cos b|{f}_{1}\rangle -\sin b|{f}_{2}\ra \quad
{\rm and} \notag \\
&& |{f}_{2}^{b}\ra =\sin b|{f}_{1}\ra +\cos b|{f}_{2}\ra,
\eeqa
where the rotation angles $a$ and $b$ are unrelated.

Next, the correlation between the polarization ($u$) and function ($f$)
degrees of freedom is given by the standard average
\beq \label{CAB1}
C(a,b) =\la {\bf e}|Z^u(a)\otimes Z^f(b)|{\bf e}\ra,
\eeq
where $Z$ is shorthand for the difference projection: $Z^u(a) \equiv
|{u}^{a}_1\ra\la{u}^{a}_1| -|{u}^{a}_2\ra\la{u}^{a}_2|$, analogous to a
$\sigma_z$ spin operation. $C(a,b)$ is thus a combination of four joint
projections such as:
\beqa \label{Pjkab}
P_{11}(a,b) &=& \la {\bf e}|\Big(|{u}_{1}^{a}\ra |f_{1}^{b}\ra \la
f_{1}^{b}|\la {u}_{1}^{a}|\Big)|{\bf e}\ra \\ \nonumber
&=& \Big|\la f_{1}^{b}|\la {u}_{1}^{a}|{\bf e}\ra \Big|^2.
\eeqa
This is all classical and all of the correlation projections $P_{jk}(a,b)$
with $j,k=1,2$ have familiar roles in classical optical polarization
theory \cite{Brosseau-Wolf}.

Gisin \cite{Gisin} observed that any quantum state entangled in the same
way as the classical pure state (\ref{efield}) will lead to violation of the
CHSH inequality, which takes the form ${\cal B} \le 2$, where
\beq \label{B}
{\cal B} = |C(a,b) - C(a',b) + C(a,b') + C(a',b')|.
\eeq
The same result will be found here, as one uses only DOF independence
and properties of positive functions and normed vectors to arrive at it (see details in Suppl. Matls. \cite{Suppl.Mat.}). We note again that the issue of entanglement itself is pertinent to the
discussion, but the usefulness of entanglement as a resource for
particular applications is not. Thus we have reached the main goal of
our theoretical background sketch. This was to demonstrate the existence
of a purely classical field theory that can exhibit a violation of the
CHSH Bell inequality.

\noindent{\bf Experimental Testing:} The remaining task is to show that experimental observation confirms this theoretical prediction, in effect shifting one's interpretation of tests of the quantum-classical border by showing that, along with quantum fields, classical fields conforming to the Shimony Bell-test criteria are capable of Bell violation. In order to make such a demonstration, a classical field source must be used. This means one producing a field that is quantum mechanical (since we believe all light fileds are intrinsically quantum), but a field whose quantum statistics are not distinguishable from classical statistics. This is only necessary up to second order in the field because the CHSH procedure engages no higher order statistics. Such sources are easily available. Since the earliest testing of laser light it has been known that a laser operated below threshold has statistical character not distinguishable from classical thermal statistics. So in our experiments we have used a broadband laser diode operated below threshold.

Our experiment repeatedly records the correlation function $C(a,b)$ defined in (\ref{CAB1}) for four different angles in order to construct the value of the Bell parameter ${\cal B}$. This is done through measurements of the joint projections $P_{jk}(a,b)$. We will describe explicitly only the recording of $P_{11}(a,b)$, identified in (\ref{Pjkab}), but the others are done similarly in an obvious way. In the classical context that we are examining, the optical field is macroscopic and correlation detection is essentially calorimetric (i.e., not requiring or employing individual photon recognition).

\noindent{\bf Polarization Tomography:} The first step is to determine tomographically the polarization state of
the test field. A polarization tomography setup is shown in Fig. 1. Using
a polarizing beam splitter and half and quarter wave plates to project
onto circular and diagonal bases, the Stokes parameters ($S_1,\ S_2,\
S_3$), relative to $S_0 = 1$, are found to be ($ -0.0827,\ -0.0920,\
-0.0158 $), providing a small non-zero degree of polarization equal to
0.125. This departure from zero requires a slight modification of
the theory presented above (see the Supplemental Material) and reduces the
maximum possible value of ${\cal B}$ able to be achieved for our specific
experimental field to ${\cal B} = 2.817$, below but close to ${\cal B} =
2\sqrt 2 = 2.828...$, the theoretical maximum for completely unpolarized
light.

\begin{figure}
\begin{center}
\includegraphics[width=6.5cm]{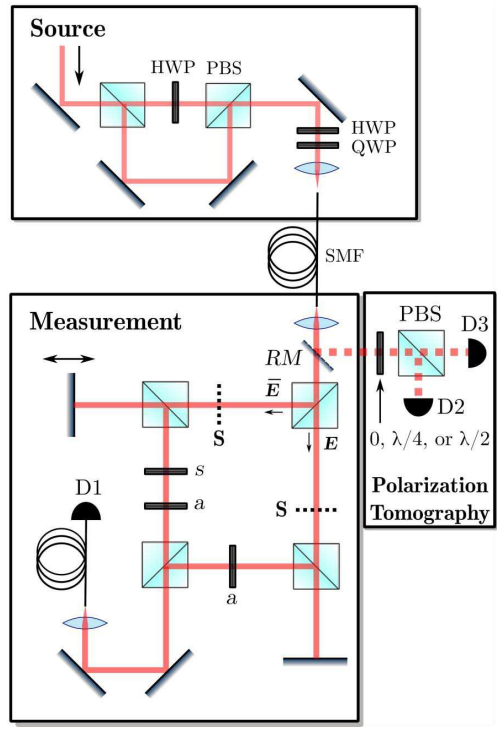}
\caption{\footnotesize The experimental setup consists of a
source of unpolarized light and a measurement using a modified
Mach-Zehnder interferometer.  Half and quarter wave plates (HWP, QWP)
control the polarization of the source.  All beam splitters are 50:50
unless marked as a polarizing beam splitter (PBS).  Intensities
needed for obtaining the required joint projections are measured at
detector D1.  Shutters S independently block arms of the
interferometer in order to measure light through the arms separately.
A removable mirror (RM) directs the light to a polarization
tomography setup, where the orthogonal components of the polarization
in the basis determined by the wave plate are measured at detectors
D2 and D3.}
\label{experiment}
\end{center}
\end{figure}

\noindent{\bf Experimental Bell Test:} The experimental test has two major components, as shown in Fig.
\ref{experiment}: a source of the light to be measured, and a
Mach-Zehnder (MZ) interferometer. The source utilizes a 780 nm laser
diode, operated in the multi-mode region below threshold, giving it a
short coherence length on the order of 1 mm. The beam is assumed statistically ergodic, stable and stationary, as commonly delivered from such a multimode below-threshold diode. It is incident on a
50:50 beam splitter and recombined on a polarizing beam splitter after
adequate delay so that the light to be studied is an incoherent mix of
horizontal and vertical polarizations before being sent to the
measurement area via a single mode fiber.  A half wave plate in one arm
controls the relative power, and thus the degree of polarization (DOP).
Quarter and half wave plates help correct for polarization changes
introduced by the fiber.

In Fig.~\ref{experiment} the partially polarized beam entering the MZ is
separated by a 50:50 beam splitter into primary test beam $|{\bf
E}\rangle$ and auxiliary beam $|\bar{\bf E}\rangle$. The two beams
inherit the same statistical properties from their mother beam and thus
both can be expressed as in Eq. (\ref{edef}), with intensities $I$ and
$\bar{I}$. The phase of the auxiliary beam $|\bar{\bf E}\rangle$ is
shifted by an unimportant factor $i$ at the beam splitter.\\

To determine the joint projection $P_{11}(a,b)$ of the test beam $|{\bf
E}\rangle$, the first step is to project the field to obtain $|{\bf
E}_1^{a}\rangle \equiv |{u}^{a}_1\ra\la{u}^{a}_1|{\bf E}\ra$. This can be
realized by the polarizer labelled ${a}$ on the bottom arm of the MZ. The
transmitted beam retains both $|f\ra$ components in function space:
\beq
|{\bf E}_{1}^{a}\ra =\sqrt{I_{1}^{a}}|u _{1}^{a}\ra (c_{11}|f_{1}^{b}\ra +
c_{12}|f_{2}^{b}\ra),
\eeq
where $I_{1}^{a}$ is the intensity, and $c_{11}$ and $c_{12}$ are
normalized amplitude coefficients with $|c_{11}|^{2}+|c_{12}|^{2} = 1$.
Here $c_{11}$ relates to $P_{11}$ in an obvious way:
$P_{11}(a,b) = I_1^a|c_{11}|^2/I$. One sees that the intensities $I$ and
$I_1^a$ can be measured directly, but not the coefficient $c_{11}$.

For $P_{11}(a,b)$ our aim is to produce a field that combines a projection
onto  $|f_1^b\ra$ in function-space with the $|u_1^{a}\ra$ projection in
polarization space. The challenge of overcoming the lack of polarizers for
projection of a non-deterministic field onto an arbitrary direction in its
independent infinite-dimensional function space is managed by a
``stripping" technique
\cite{Suppl.Mat.} applied to the auxiliary ${\bf \bar E}$ field in the
left arm. We pass ${\bf \bar E}$ through a polarizer rotated from the
initial $|u_{1}\ra - |u_{2}\ra$ basis by a specially chosen angle $s$, so
that the statistical component $|f_{2}^{b}\ra$ is stripped off. The
transmitted beam $|{\bf\bar E}_{1}^s\ra$ then has only the $|f_{1}^{b}\ra$
component, as desired: $|{\bf \bar E}_{1}^s\ra = i\sqrt{
\bar{I}_{1}^{s}}|{u}_{1}^{s}\ra |f_{1}^{b}\ra$. Here $\bar{I}_{1}^{s}$ is
the corresponding intensity and the special stripping angle $s$ is given
by $\tan s = (\kappa_{1}/\kappa_{2})\tan b$ (see \cite{Suppl.Mat.,
Qian-Eberly-arX13}).

The function-space-oriented beam $|{\bf \bar E}_{1}^s\ra$ is then sent
through another polarizer ${a}$ to become $|{\bf \bar E}_{1}^{a}\ra
=|{u}_{1}^{a}\ra \la {u}_{1}^{a}|{\bf \bar E}_{1}^{a}\ra =
i\sqrt{\bar{I}_1^{a}}|{u}_{1}^{a}\ra |f_{1}^{b}\ra$, where
$\bar{I}_{1}^{a}$ is the corresponding intensity. Finally, the beams
$|{\bf E}_{1}^{a}\ra$ and $|{\bf \bar E}_{1}^{a}\ra$ are combined by a
50:50 beam splitter which yields the outcome beam $|{\bf E}_{1}^{T}\ra =
(|\bar{\bf E}_{1}^{a}\ra + i|{\bf E}_{1}^{a}\ra)/\sqrt{2}$. The total
intensity $I_{1}^{T}$ of this outcome beam can be easily expressed in
terms of the needed coefficient $c_{11}$.

Some simple arithmetic will immediately provide the joint projection
$P_{11}(a,b)$ in terms of various measurable intensities:
\beq \label{eq:prob}
P_{11}(a,b) =(2I_{1}^{T} - \bar{I}_{1}^{a} -
I_{1}^{a})^{2}/4I\bar{I}_{1}^{a}.
\eeq
Other $P_{jk}(a,b)$ values can be obtained similarly by rotations of
polarizers $a$ and $s$. To make our measurements, polarizers $a$ were
simultaneously rotated using motorized mounts, while the third polarizer
$s$ was fixed at different values in a sequence of runs.

\noindent{\bf Results:} For each angle, measurements were made at detector D1 for the total
intensity $I^T$, and the separate intensities from each arm $I^{a}$ and
$\bar{I}^{a}$ by closing the shutters $S$ alternately. In this way, the
measurements of the polarization space and statistical amplitude
space are carried out separately. From these
measurements the needed correlations $C(a,b)$ were determined and Eq.
\eqref{B} used to evaluate the CHSH parameter $\cal B$.

\begin{figure}[t!]
\begin{center}
\includegraphics[width=7.6cm]{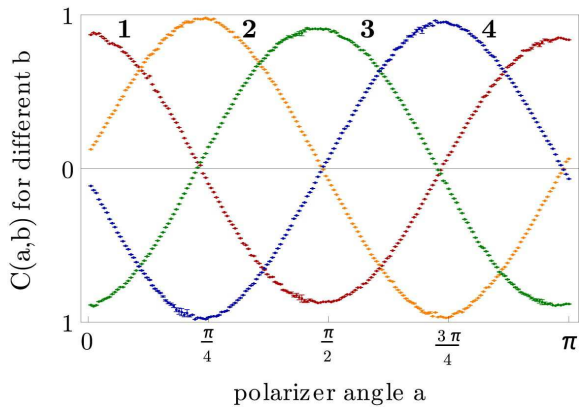}
\caption{\footnotesize Plots of the correlation functions C(a,b) obtained by rotating polarizer $a$ in the polarization space and holding angle $b$ in the function space constant. Curves 1 to 4 correspond to different fixed values of $b$ separated by $\pi/4$. The invariant cosine function required to violate the Bell inequality is clearly present. Error bars are included but scarcely visible.}
\label{fig:results}
\end{center}
\end{figure}

Fig.~\ref{fig:results} shows $C(a,b)$ obtained by measuring the  joint
projections $P_{jk}(a,b)$ for a complete rotation of polarizer $a$, with
different curves corresponding to $b$ (and thus $s$) fixed at different
values. It is apparent from the near-identity of the curves that, to good
approximation, the correlations are a function of the difference in
angles, i.e. $C(a,b)=C(a-b)$. The maximum value for
$\cal B$ can then be found straightforwardly from any one of the curves in
Fig.
~\ref{fig:results}. Among them the smallest and largest values of $\cal B$
(obtained for curves 1 and 4), are $2.548 \pm 0.004$
and $2.679 \pm 0.007$.\\

To be careful, we note that in our experiments the field was almost but
not quite completely unpolarized, thus not quite the same field sketched
in the Background Theory paragraphs. Thus we couldn't expect to get the
maximum quantum result ${\cal B} = 2\sqrt 2 = 2.828...$ for the Bell
parameter, but the values achieved also present a strong violation. The
background theory is mildly more complicated for partially polarized
rather than unpolarized light, but when worked out for the degree of
polarization of our light beams (see \cite{Qian-Eberly-arX13} and the Suppl. Matl. \cite{Suppl.Mat.}) it supports the values we observed.

\noindent{\bf Summary:} In summary, we first sketched the purely classical theory of optical beam fields (\ref{Efield}) that satisfy the Bell-test criteria of Shimony \cite{Shimony, Suppl.Mat.}. Their bipartite pure state form shows the entanglement of their two independent degrees of freedom \cite{Gerry-Dowling}. The classical theory defines them as dynamically probabilistic fields, meaning that individual field measurements yield values that cannot be predicted except in an average sense, which is another feature shared with quantum systems but also associated for more than 50 years with the well-understood and well-tested optical theory of partial coherence \cite{Brosseau-Wolf}. Our theoretical sketch for the simplest case, unpolarized light, indicated that such fields or states are predicted to possess a range of correlation strengths equal to that of two-party quantum systems, that is, outside the bound  ${\cal B} \le 2$ of the CHSH Bell inequality and potentially as great as ${\cal B} = 2\sqrt 2$. In our experimental test we used light whose statistical behavior (field second-order statistics) is indistinguishable from classical, viz., the light from a broadband laser diode operating below threshold. Our detections of whole-beam intensity are free of the heralding requirements familiar in paired-photon CHSH experiments. Repeated tests confirmed that such a field can strongly violate the CHSH Bell inequality and can attain Bell-violating levels of correlation similar to those found in tests of maximally entangled quantum systems.

One naturally asks, how are these results possible? We know that a field with classically random statistics is a local real field, and we also know that Bell inequalities prevent local physics from containing correlations as strong as what quantum states provide. But the experimental results directly contradict this. The resolution of the apparent contradiction is not complicated but does mandate a shift in the conventional understanding of the role of Bell inequalities, particularly as markers of a classical-quantum border. Bell himself came close to addressing this point. He pointed out \cite{BellSci} that even adding classical indeterminism still wouldn't be enough for any type of hidden variable system to overcome the restriction imposed by his inequalities. This is correct as far as it goes but fails to engage the point that local fields can be statistically classical and exhibit entanglement at the same time. For the fields under study, the entanglement is a strong correlation that is intrinsically present between the amplitude and polarization degrees of freedom, and it is embedded in the field from the start (as it also is embedded {\em ab initio} in any quantum states that violate a Bell inequality). The possibility of such pre-existing structural correlation is bypassed in a CHSH derivation. Thus one sees that Bell violation has less to do with quantum theory than previously thought, but everything to do with entanglement.

\noindent{\bf Funding Information:} NSF PHY-0855701, NSF PHY-1203931, and DARPA DSO Grant No.
W31P4Q-12-1-0015.

\noindent{\bf Acknowledgments:} The authors acknowledge helpful
discussions with many colleagues, particularly including M. Lewenstein,
as well as A.F. Abouraddy, A. Aspect, C.J. Broadbent, L. Davidovich,  E.
Giacobino, G. Howland, A.N. Jordan, P.W. Milonni, R.J.C. Spreeuw, and
A.N. Vamivakas.

\newpage
\newpage

\section{Supplementary Material}

\noindent{\bf Shimony's Bell-Test Conditions:} In his extended analyses
\cite{Shimony} of Bell inequalities and their testing, Shimony recognized
that in order to deserve serious attention, an alternative non-quantum
theory entering what is considered a quantum domain (in our case, the
domain of Bell test violation) needs to embrace in some way aspects of
Nature that appear completely random, i.e., purely probabilistic. These
aspects are dealt with by quantum theory in well known ways, and this is
the reason Bell once raised the issue of classical indeterminism
\cite{BellSci}, but without going as far as classical entanglement.
Shimony summarized these considerations by naming three key features, all
of which should be considered, in his words, ``... as established parts of
physical theory: (I) In any state of a physical system $S$ there are some
eventualities which have indefinite truth values. (II) If an operation is
performed which forces an eventuality with indefinite truth value to
achieve definiteness ... the outcome is a matter of chance. (III) There
are `entangled systems' (in Schr\"odinger's phrase) which have the
property that they constitute a composite system in a pure state, while
neither of them separately is in a pure state."  Here by eventualities
Shimony means measurement outcomes. We have identified the electric field
as it is dealt with in the standard classical theory of partial coherence
and partial polarization as a physical system satisfying all three
conditions.

\noindent{\bf Partially Polarized Fields:} Classical statistical light of any
degree of polarization can be treated exactly as the unpolarized light in
the text, with a small change. One needs to insert the parameters that
allow the two orthogonal components provided by the Schmidt decomposition
to have different intensities. These parameters, $\ka_1$ and $\ka_2$, are
the (real) Schmidt eigenvalues. They satisfy $\ka_1^2 + \ka_2^2 = 1$, and
both equal $1/\sqrt 2$ in the completely unpolarized case treated in the
text. Any intensity-normalized partially polarized field can then be
written as in text Eqn. (3), but with $\ka_1$ and $\ka_2$ attached
\cite{Qian-Eberly-arX13}:
\beq \label{e-def}
|{\bf e}\ra = \kappa_{1}|u_1\ra |f_1\ra + \kappa_{2}|u_2\ra |f_2\ra.
\eeq
It is clear that the field is entangled between polarization and amplitude
unless one of the $\ka$'s is zero, in which case $|{\bf e}\ra$ is plainly
separable. All of the formulas for unpolarized light will change, but only
to the extent that the presence of the $\ka$'s requires. The conventional
degree of polarization $P$ is given in terms of the $\ka$'s by $P =
|\ka_1^2 - \ka_2^2|$ \cite{Qian-Eberly-OL}, and this can be used to find
the Schmidt coefficients. For the experimental classical statistical optical light field we
obtained $\kappa_{1},\ \ka_{2} = 0.750,\ \ 0.661.$  Another interesting
formula is the full result for the partially polarized Bell parameter:
\beqa \label{B+kappas}
{\cal B} &=& \cos 2a(\cos 2b - \cos 2b') \nonumber \\
&+& \cos 2a'(\cos 2b + \cos 2b') \nonumber \\
&+& 2\ka_1\ka_2\{\sin 2a(\sin 2b - \sin 2b') \nonumber \\
&+& \sin 2a'(\sin 2b + \sin 2b') \}.
\eeqa
One sees two uncomplicated limits: when either $\ka$ is zero (no
entanglement) no result higher than ${\cal B} = 2$ can be achieved, and
when $\ka_1 = \ka_2 = 1/\sqrt 2$ (maximal entanglement) the Tsirelson
bound can be reached: ${\cal B} = 2\sqrt 2$. Also, if one follows Gisin's approach \cite{Gisin} by choosing the rotation angles as $a=0$, $a'=\pi/4$, and $cos2b=-cos2b'=1/\sqrt{1+4|\kappa_1\kappa_2|^2}$, the Bell parameter becomes
\begin{equation}
{\cal B} = 2/\sqrt{1+4|\kappa_1\kappa_2|^2}.
\end{equation}
Apparently, there will be a Bell violation (${\cal B}$>2) as long as $\kappa_1\kappa_2\neq 0$, i.e., when there is non-zero entanglement.

\noindent{\bf Basis stripping, rotation, and projection in amplitude function space:} To observe correlation, it is essential to be able to access and measure both polarization and amplitude function degrees of freedom. Unfortunately, unlike the polarization degree of freedom, there is no systematic technology working directly in the infinite-dimensional amplitude function
space that can project a non-deterministic field onto an
arbitrary basis $|f_{1}^{b}\ra$ in that space. This requires an innovation using an
indirect measurement such as incorporated in the experimental setup
sketched in Fig. 1 of the text. In our setup we employ
an auxiliary beam that contains only the $|f_{1}^{b}\ra$ basis to
interfere with the primary test beam (which in general contains both
$|f_{1}^{b}\ra$ and $|f_{2}^{b}\ra$). By interference measurements one is able to obtain
the $|f_{1}^{b}\ra$ information from the test beam, and finally determine
the needed $c_{11}$ value given in equation (9) of the main text.

This section shows specifically how to ``strip" off and rotate a basis in the statistical amplitude function space of the auxiliary field $|\bar{\bf E}\rangle$, which shares the same statistical properties as the primary test beam. For generality, we take the auxiliary beam as initially in the form of Eq.~(\ref{e-def}) with arbitrary $\kappa_{1}, \kappa_{2}$. Such a beam can always be rewritten in the rotated amplitude function space basis
$|f_{1}^{b}\ra$, $|f_{2}^{b}\ra$, i.e,
\begin{eqnarray}
|\bar{\bf E}\rangle
&=&\sqrt{\bar {I}}\Big[\left(\kappa _{1}\cos b|u_{1}\rangle -\kappa _{2}\sin
b|u_{2}\ra \right)
|f_{1}^{b}\rangle \notag \\
&& + \left(\kappa _{1}\sin b|u_{1}\rangle +\kappa _{2}\cos
b|u_{2}\ra \right)|f_{2}^{b}\rangle \Big].  \label{before strip}
\end{eqnarray}
One notes from the second term of the equation that a properly
chosen polarizer that blocks completely the polarization-space component
$\kappa
_{1}\sin b|u_{1}\rangle +\kappa _{2}\cos b|u_{2}\rangle $ will effectively
strip off the amplitude function space basis vector $|f_{2}^{b}\rangle$.

Such a stripping polarizer $|{u}_{1}^{s}\ra$ is oriented with respect
to the polarization space
basis $|{u}_{1}\ra$,\ $|u_2\ra$ by a specific angle we have called $s$, i.e.,
\begin{equation}
|{u}_{1}^{s}\ra =\cos s|{u}_{1}\ra -\sin s|{u}_{2}\ra.
\end{equation}
Then the stripping condition is directly given as
\begin{equation}
\la {u}_{1}^{s}|\left(\kappa _{1}\sin b|u_{1}\rangle +\kappa _{2}\cos
b|u_{2}\ra\right) = 0,
\end{equation}
which specifies the rotation angle $s$ according to
\begin{equation}
\tan s=(\kappa _{1}/\kappa _{2})\tan b, \label{stripping condition}
\end{equation}
so $s$ is determined by the values of $\ka_{1}$ and $\ka_2$ for any choice of
rotation angle $b$ in the amplitude function space.

As a result of this stripping polarizer, the beam (\ref{before strip})
becomes
\begin{eqnarray}
&&|\bar{\bf E}_{1}^{s}\rangle =|{u}_{1}^{s}\rangle \langle {u}_{1}^{s}|
\bar{\bf E}\rangle \notag \\
&=&\sqrt{\bar{I}}(\kappa _{1}\cos b\cos s+\kappa _{2}\sin b\sin s)|{u}
_{1}^{s}\rangle |{f}_{1}^{b}\rangle, \label{after strip}
\end{eqnarray}
where the amplitude function space component $|f_{2}^{b}\rangle$ of the
transmitted beam is completely stripped off.

One notes immediately that arbitrary rotations of the function space basis $|f_{1}^{b}\ra$ (i.e., arbitrary change of angle $b$) for the auxiliary beam can be effectively realized by the change of angle $s$ through Eq. (\ref{stripping condition}). Consequently, as described in the main text, by further interference of this auxiliary beam with the primary test beam, one is able to project the test beam onto any function space basis $|f_{1}^{b}\ra$ with an arbitrary angle $b$. This is exactly how we make measurements in the amplitude function space of the statistical light beam.


\begin{thebibliography}{99}
\bibitem{circpol} Elliptical polarization needs different notation, but
the results are the same.
\end{thebibliography}
\end{document}